\documentclass[10pt,conference]{IEEEtran}
\usepackage[numbers,sort&compress]{natbib}
\usepackage[T1]{fontenc}
\usepackage{booktabs}
\usepackage{csquotes}
\usepackage{paralist}
\usepackage{pifont}
\usepackage{makecell}
\usepackage{siunitx}
\usepackage{subcaption}
\usepackage{threeparttable}
\usepackage{xspace}
\newcommand{\cmark}{\ding{51}}%
\newcommand{\xmark}{\ding{55}}%
\usepackage{listings}
\usepackage{amsmath}
\usepackage{makecell}
\usepackage{tikz}
\usepackage{multirow}
\usepackage{textcomp}
\usepackage[htt]{hyphenat}
\usepackage{xcolor,tcolorbox}
\usepackage{microtype}
\usetikzlibrary{decorations.pathmorphing}
\makeatletter
\newenvironment{btHighlight}[1][]
{\begingroup\tikzset{bt@Highlight@par/.style={#1}}\begin{lrbox}{\@tempboxa}}
{\end{lrbox}\bt@HL@box[bt@Highlight@par]{\@tempboxa}\endgroup}

\usepackage{enumitem}

\newcommand\percentage[2][round-precision = 2]{% default precision: 2
    \SI[round-mode = places,
        scientific-notation = fixed, fixed-exponent = 0,
        output-decimal-marker={.}, #1]{#2e2}{\percent}%
}

\newcommand\percentageint[2][round-precision = 0]{%
    \SI[round-mode = places,
        scientific-notation = fixed, fixed-exponent = 0,
        output-decimal-marker={.}, #1]{#2e2}{\percent}%
}

\newcommand{\etal}[0]{\emph{et al.}\xspace}

\newlist{questions}{enumerate}{2}
\setlist[questions,1]{label=RQ\arabic*.,ref=RQ\arabic*}
\setlist[questions,2]{label=\thequestionsi.{\arabic*.},ref=\thequestionsi(\arabic*)}

\definecolor{prettyblue}{HTML}{3465a4}
\definecolor{prettyyellow}{HTML}{fce94f}

\definecolor{bugcolor}{HTML}{cc0000}
\definecolor{scannercolor}{HTML}{F57900}
\definecolor{repaircolor}{HTML}{AD7FA8}

\newcommand\btHL[1][]{%
  \begin{btHighlight}[#1]\bgroup\aftergroup\bt@HL@endenv%
}
\def\bt@HL@endenv{%
  \end{btHighlight}%
  \egroup
}
\newcommand{\bt@HL@box}[2][]{%
  \tikz[#1]{%
    \pgfpathrectangle{\pgfpoint{1pt}{0pt}}{\pgfpoint{\wd #2}{\ht #2}}%
    \pgfusepath{use as bounding box}%
    \node[anchor=base west, fill=orange!30,outer sep=0pt,inner xsep=1pt, inner ysep=0pt, rounded corners=3pt, minimum height=\ht\strutbox+1pt,#1]{\raisebox{1pt}{\strut}\strut\usebox{#2}};
  }%
}
\makeatother

\lstdefinestyle{patch}{
    basicstyle=\ttfamily\fontsize{6}{6}\selectfont,
    moredelim=**[is][{\btHL[fill=green!30]}]{`}{`},
    moredelim=**[is][{\btHL[fill=red!30]}]{@}{@},
    moredelim=**[is][{\btHL[fill=scannercolor!30]}]{~~}{~~},
    moredelim=**[is][{\btHL[fill=prettyblue!30]}]{??}{??},    
    escapeinside={(*}{*)},
    numbers=left,
    keywordstyle=\ttfamily\bfseries,
    stringstyle=\ttfamily\itshape
}

\definecolor{resultboxgray}{gray}{0.98}

\newcommand{\secref}[1]{Section~\ref{#1}}

\tcbuselibrary{skins}
\tcbset{shield externalize} 

\newtcolorbox{answerbox}[2][]{
    blanker,
    left=3mm,
    right=3mm,
    borderline west={1.2pt}{0pt}{black},
    title={#2},
    fonttitle=\bfseries,
    coltitle=black,
    #1}

\AtBeginDocument{%
  \providecommand\BibTeX{{%
    \normalfont B\kern-0.5em{\scshape i\kern-0.25em b}\kern-0.8em\TeX}}}

\usepackage{hyperref}

\begin{document}

\title{Extracting Fix Ingredients using Language Models}

%\author{\IEEEauthorblockN{Anonymous Authors}}
\author{\IEEEauthorblockN{Julian Aron Prenner}
\IEEEauthorblockA{\textit{Free University of Bozen-Bolzano}\\
Bozen-Bolzano, Italy \\
prenner@inf.unibz.it}
\and
\IEEEauthorblockN{Romain Robbes}
\IEEEauthorblockA{\textit{Univ. Bordeaux, CNRS, Bordeaux INP},\\ \textit{LaBRI, UMR 5800} \\
Bordeaux, France \\
romain.robbes@u-bordeaux.fr}
}

\def\rqiDfjCoverIT{93\%}
\def\rqiDfjCoverIF{77\%}
\def\rqiDfjCoverIC{61\%}
\def\rqiDfjCoverIM{55\%}

\def\rqiTssbCoverIT{91\%}
\def\rqiTssbCoverIF{69\%}
\def\rqiTssbCoverIC{49\%}
\def\rqiTssbCoverIM{37\%}
\def\rqiDfjRatioWithNoIngredient{15\%}
\def\rqiTssbRatioWithNoIngredient{56\%}
\def\rqiDfjMedianFixIngrCount{4}
\def\rqiTssbMedianFixIngrCount{0}

\newcommand{\xmarkrect}{\tikz[baseline=(X.base)]\node[fill=red!30,rectangle,inner sep=2pt, outer sep=2pt, rounded corners=2pt] (X) {\xmark};}
\newcommand{\cmarkrect}{\tikz[baseline=(X.base)]\node[fill=green!30,rectangle,inner sep=2pt, outer sep=2pt, rounded corners=2pt] (X) {\cmark};}

\newcommand{\xmarkrectt}[1]{\tikz[baseline=(X.base)]\node[fill=red!30,rectangle,inner sep=2pt, outer sep=2pt, rounded corners=2pt] (X) {\xmark\textsubscript{#1}};}
\newcommand{\cmarkrectt}[1]{\tikz[baseline=(X.base)]\node[fill=green!30,rectangle,inner sep=2pt, outer sep=2pt, rounded corners=2pt] (X) {\cmark\textsubscript{#1}};}

\newcommand{\myparagraph}[1]{\noindent \textbf{#1}}
\maketitle

\begin{abstract}
Deep learning and language models are increasingly dominating automated program repair research. While previous generate-and-validate approaches were able to find and use fix ingredients on a file or even project level, neural language models are limited to the code that fits their input window. In this work we investigate how important identifier ingredients are in neural program repair and present ScanFix, an approach that leverages an additional \emph{scanner} model to extract  identifiers from a bug's file and potentially project-level context. We find that lack of knowledge of far-away identifiers is an important cause of failed repairs.
Augmenting repair model input with scanner-extracted identifiers yields relative improvements of up to 31\%. However, ScanFix is outperformed by a model with a large input window (> 5k tokens). When passing ingredients from the ground-truth fix, improvements are even higher. This shows that, with refined extraction techniques, ingredient scanning, similar to fix candidate ranking, could have the potential to become an important \enquote{subtask} of future automated repair systems. At the same time, it also demonstrates that this idea is subject to Sutton's bitter lesson and may be rendered unnecessary by new code models with ever-increasing context windows.

\end{abstract}


\section{Introduction}
Neural program repair methods (NPR), in particular those relying on large pre-trained language models, are increasingly replacing previous automated program repair (APR) approaches. On multiple benchmarks, NPR methods outperform traditional generate-and-validate (G\&V) methods by a wide margin~\citep{lutellierCoCoNuTCombiningContextaware2020, prennerCanOpenAICodex2022a, jiangImpactCodeLanguage2023}.

Before adopting neural techniques, the field of APR strongly relied on the plastic surgery hypothesis~\citep{barrPlasticSurgeryHypothesis2014}. The plastic surgery hypothesis states that a correct fix can be crafted from code elements (so-called ingredients or donor code) already found in a project's code base. Consequently, many search-based APR tools operate at the project level. Here, the search for useful repair ingredients spans the entire projects or beyond (e.g., code corpora).

Recent neural repair models, on the other hand, mostly operate on the level of code snippets. These snippets are usually comprised of the buggy code (i.e., bug location) and code surrounding the bug (local context). This local context can be the surrounding method (i.e., the method under repair) or a predefined number of code lines. Unfortunately, this local context is not very large (often a few dozens of lines of code~\citep{prenner2023out}). This means that repair-relevant project-level information is often not to be found in the model input: it is \emph{out of context}. For instance, if a fix requires calling a specific method declared at a different location in the project or even in the same class or file, the model may generate an invalid patch by guessing the wrong method name.     

\begin{figure}[htbp]
\centering
\includegraphics[width=.7\columnwidth]{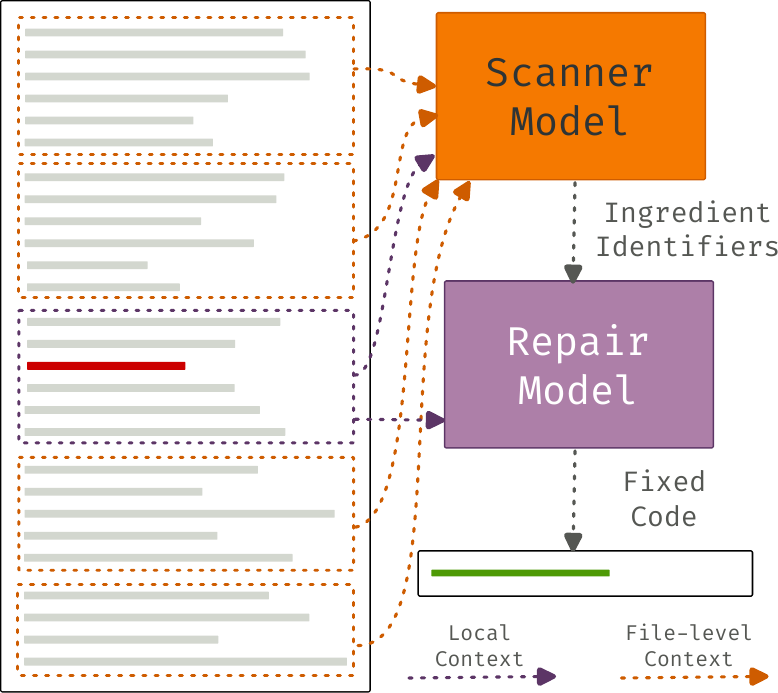}
\caption{ScanFix during inference: the {\btHL[fill=scannercolor!50]scanner model} receives the {\btHL[fill=bugcolor!50]buggy code location} with a local context (e.g. 10 lines before and after) as well as code snippets from the file under repair from which it extracts relevant identifier ingredients. These identifiers are passed on to the actual {\btHL[fill=repaircolor!50]repair model}.
}
\label{fig:scan-fix-overview}
\end{figure}

\myparagraph{Motivating Example.}  
To further illustrate this problem, we turn to a motivating example. Consider bug Chart\#10 from Defects4J~\citep{justDefects4JDatabaseExisting2014}. This is a bug in the commonly used JFreeChart library \cite{jFreeChart}. 
The parameter (\texttt{toolTipText}) passed to the method under repair must be properly escaped using the \emph{already existing} method \texttt{htmlEscape} defined inside the class \texttt{ImageMapUtilities}.

\begin{figure}[htbp]
\centering
\begin{lstlisting}[language=Java,style=patch,numbers=none,xleftmargin=0pt,xrightmargin=0pt,framesep=0pt]
public String generateToolTipFragment(String toolTipText) {
    return " title=\"" 
        + `ImageMapUtilities.htmlEscape(`toolTipText`)`
        + "\" alt=\"\"";
}
\end{lstlisting}
\caption{Bug Chart\#10 from Defects4J with ground-truth fix.}
\label{fig:d4j-chart10-patch}
\end{figure}

We challenge ChatGPT (GPT-3.5) to fix this bug by prompting it with the buggy method, along with its leading comment, the corresponding error message and an appropriate prompt command (\enquote{Fix a bug in the following code}). ChatGPT proposes to use the \texttt{StringEscapeUtils.escapeHtml} method from the Apache Commons library. However, JFreeChart does not depend on Apache Commons, thus the in and of itself correctly generated import would cause a compilation error.
If, on the other hand, we force ChatGPT to make use of the identifier ingredients \texttt{ImageMapUtilities} and \texttt{escapeHtml} it will output a correct patch.
These \emph{identifier ingredients}, and how to leverage them effectively, are the central theme of this work. Identifier ingredients may include variable names, method and function names as well as names of constants, keyword arguments, classes and interfaces.

\myparagraph{Previous work.}
Previous work has already explored simple means of augmenting model input with ingredient information. These approaches where either \enquote{static} (e.g., augmenting model input with instance variable declarations from the class under repair~\citep{chenSequenceRSequencetoSequenceLearning2021, yeNeuralProgramRepair2022}) or fairly basic (extracting identifiers using a lexical similarity measure~\citep{xiaRevisitingPlasticSurgery2023}). Section~\ref{sec:related-work} gives a more detailed discussion of related work.

\myparagraph{This work.}
We expand on previous work by
\begin{inparaenum}[(I)]
    \item providing an analysis of identifier ingredients on two benchmarks/datasets and
    \item using a Transformer-based neural network to extract identifier ingredients from the file or project under repair.
\end{inparaenum}
We answer the following research questions:
\begin{questions}
\item \emph{How important are identifier ingredients in program repair? (\secref{sec:rq1})} Analyzing an APR benchmark (Defects4J~\citep{justDefects4JDatabaseExisting2014}) and a APR dataset (TSSB-3M~\citep{richterTSSB3MMiningSingle2022}) we explore how common identifier ingredients are, where they can be found (e.g., in the input window, the file or project under repair) and how often they are not part of a typical Transformer-based repair model's input.
\item \emph{How do identifier ingredients impact repair success? (\secref{sec:rq2})} Building on RQ1, we investigate how repair success of NPR repair models is affected by identifier ingredients, in particular when they fall outside the local-context (i.e., the input window).
\item \emph{How effective can specially trained models extract identifiers from the file-level context? (\secref{sec:rq3})}
We train a specific ingredient-extraction model (scanner model). We then evaluate whether our scanner can extract fix ingredients from a larger context on a large dataset of bugs with file-level context based on TSSB-3M.
\item \emph{How well can NPR models leverage ingredient information? (\secref{sec:rq4})}
We present ScanFix: it combines the scanner model from RQ3 with a Transformer-based repair model. We evaluate ScanFix and compare its performance with several baselines.
\end{questions}

\myparagraph{Findings.} We find that fix ingredients are commonly required for repair (RQ1) and an important factor for repair success (RQ2). Our Transformer-based scanner model is able to extract identifier ingredients with an F1 score of 27\% on a dataset of 85'000 Python bugs (RQ3). Finally, augmenting the repair model input 
with ingredients extracted by the scanner model (ScanFix) yields modest relative improvement between 7\% and 31\%. While this is considerably better than our naive baseline that randomly selects identifier ingredients, ScanFix remains behind a baseline with very large context and no ingredient augmentation (RQ4).

\myparagraph{Implications.} Our findings show that extracting identifiers from a wider context is generally viable. However, a baseline that is endowed with a very large input window doing \emph{no} ingredient augmentation outperforms ScanFix.
Interestingly, augmenting with the \enquote{correct} ingredients from the ground-truth fix leads to even better results (by a wide margin). This shows that the idea behind ScanFix has potential if the accuracy of ingredient extraction can be further improved. We discuss more implications, and the limitations of this work, in Section~\ref{sec:implications}.
Code and scripts used in this work are provided online\cite{scanfixReplication}.
\section{Related Work}
\label{sec:related-work}

The significance of ingredient code is widely known in the APR research community. Barr \etal \cite{barrPlasticSurgeryHypothesis2014} show compelling evidence that code changes introduced in a commit can be grafted from the parent commit (plastic surgery hypothesis). Similar findings are reported by Martinez \etal \cite{martinezFixIngredientsAlready2014} who found that over half of the studied commits were composed of already existing code tokens (redundancy assumption). These insights are heavily exploited by search-based APR techniques~\citep{gouesGenProgGenericMethod2012,yuanARJAAutomatedRepair2020,  martinezUltraLargeRepairSearch2018}, that try to construct a patch from existing donor code found in their search space (e.g., the project under repair).

The plastic surgery hypothesis, in a way, was already reflected in very early NPR research. Chen \etal \cite{chenSequenceRSequencetoSequenceLearning2021} use an LSTM-based repair model to fix bugs in short Java methods (SequenceR). Beyond the method under repair, they augment model input with field variable declarations and signatures of methods and constructors of the class under repair. This idea was also adopted in later work (e.g., RewardRepair~\citep{yeNeuralProgramRepair2022}).

While SequenceR's approach is fully static, Xia \etal \cite{xiaRevisitingPlasticSurgery2023} propose a more dynamic approach which they call \enquote{relevant-identifier prompting}. They use a Levenshtein-based similarity measure to extract possibly useful identifier ingredients from the file under repair. Extracted identifier ingredients are further filtered by discarding any non-accessible symbols by means of static analysis. These identifiers are incorporated into an LLM's \enquote{repair prompt}.

A more general approach is Retrieval-Augmented Generation (RAG), where a database is queried for relevant code fragments to include in an LLM's prompt~\cite{lewis2020retrieval}. This approach is used, for example, in RAP-GEN~\cite{wangRAPGenRetrievalAugmentedPatch2023a}, \textsc{SarGam}~\citep{liuAutomatedCodeEditing2024}, AutoCodeRover~\citep{zhangAutoCodeRoverAutonomousProgram2024} and RepairAgent~\citep{bouzeniaRepairAgentAutonomousLLMBased2024}. In contrast, the method proposed in this work aims to be \enquote{space-efficient} and require as little input window space as possible. Hence, instead of including entire code fragments we work with relevant, pre-filtered identifiers.

An alternative approach to integrate project-level ingredients into the repair process is to train the repair model on project code as done by SelfAPR~\citep{yeSelfAPRSelfsupervisedProgram2023} (\enquote{project-specific training}) or FitRepair~\citep{xiaRevisitingPlasticSurgery2023} (\enquote{knowledge-intensified fine-tuning}). In this way, information about identifiers and other ingredients is directly encoded into the models weights. Because such systems require a training step for each project before the repair process can start, their training overhead is rather high.

For Defects4J (v1.0), Yang \etal \cite{yangWhereWereRepair2021} provide an extensive analysis of donor code origin. Their notion of donor code includes arbitrary code fragments, such as statements, operators, literals, or even code blocks. 

\myparagraph{This work.}
Similar to Xia \etal \cite{xiaRevisitingPlasticSurgery2023}, we focus on identifier ingredients. However, we use a more sophisticated technique to extract identifiers from code: we employ a dedicated \enquote{scanner} neural network to extract relevant identifiers from file-level or project-level code. Moreover, unlike \enquote{relevant-identifier prompting}, we fine-tune a code LLM to learn to exploit the identifiers found by the scanner module. Finally, we also provide a study of identifier ingredients, analyzing over 800,000 bugs. This study is different from the one by Yang \etal \cite{yangWhereWereRepair2021} in three important aspects: \begin{inparaenum}[(I)]
\item we focus on identifier ingredients (as opposed to code fragments),
\item we focus on NPR (as opposed to traditional search-based APR),
\item we include version 2 of the Defects4J benchmark and a second much larger bug dataset (TSSB-3M).
\end{inparaenum}

\section{Identifier Ingredients (RQ1)}\label{sec:rq1}

\newcommand{\bugIds}[0]{$bug_{ids}$\xspace}
\newcommand{\fixIds}[0]{$fix_{ids}$\xspace}
\newcommand{\fixAll}[0]{$fix_{all}$\xspace}
\newcommand{\winIds}[0]{$win_{ids}$\xspace}
\newcommand{\winOut}[0]{$win_{out}$\xspace}
\newcommand{\winIn}[0]{$win_{in}$\xspace}
\newcommand{\fileIds}[0]{$file_{ids}$\xspace}
\newcommand{\fileOut}[0]{$file_{out}$\xspace}
\newcommand{\fileIn}[0]{$file_{in}$\xspace}
\newcommand{\methodIds}[0]{$mth_{ids}$\xspace}
\newcommand{\methodOut}[0]{$mth_{out}$\xspace}
\newcommand{\methodIn}[0]{$mth_{in}$\xspace}
\newcommand{\projectIds}[0]{$proj_{ids}$\xspace}
\newcommand{\projectOut}[0]{$proj_{out}$\xspace}
\newcommand{\projectIn}[0]{$proj_{in}$\xspace}

For our analysis, we rely on two commonly used bug datasets.
First, TSSB-3M~\citep{richterTSSB3MMiningSingle2022}, a collection of over three million single statement Python bugs, chosen to increase robustness of the results due to its size. %and thus  Size is the main reason for choosing TSSB-3M, which is one of the largest collections of bug fixing commits to date.  Its considerable size of three million bugs warrants robust results.
Second, Defects4J 2.0, a dataset of 835 bugs in large Java projects~\citep{justDefects4JDatabaseExisting2014}. Unlike TSSB-3M, Defects4J provides full project code for each bug. We find that identifier ingredients are prevalent, with a large proportion likely out of context for a typical repair model.

\subsection{Mining and Pre-Processing}
We first mine buggy/fixed file pairs from GitHub and apply minor pre-processing, as described next. % such as breaking up multi-line strings and comments as well as removing duplicates. These steps are described below.

\myparagraph{Mining.}
Bugs in TSSB-3M are single statement bugs and thus affect a single file only. Each bug is provided as a short diff with limited local context (< 10 lines). To obtain full file-level context we use the commit hashes provided by TSSB-3M and the GitHub API to download the complete file contents before and after the fixing commit. Due to API call limitations, this step takes several weeks. % (with three API tokens).
While Defects4J is small in terms of numbers of bugs (< 1000), each bug comes with full project-level context; no mining is necessary.  

\myparagraph{Pre-processing.}
Comments and strings (docstring) can span multiple lines in Python and Java. In order to be able to extract identifiers, we would like any \enquote{line-boundary} snippet to be \enquote{lexable} (i.e., accepted by a Python or Java lexer). We use  TreeSitter \cite{treesitter} to break up any multi-line strings, docstrings or comments into sequences of single line strings/comments. % respectively.

\myparagraph{Deduplication.}\label{sec:dedup}
Even though deduplication measures have been taken in the creation of TSSB-3M~\citep{richterTSSB3MMiningSingle2022}, we found that it contains a significant number of duplicates. We use the commit hash of the fixing commit as a deduplication key. This reduces the number of bugs from over 3 million to a bit over 900,000.

\myparagraph{Filtering.}
For three Defects4J bugs (\#25, \#42 and \#62 in the Lang project), we run into encoding issues (all three bugs are related to HTML unity escaping and unicode encoding). These three bugs have been omitted from all of our experiments.

\myparagraph{Local Context.}
Previous work found that the size of local context can have considerable impact on repair performance~\citep{prenner2023out}. 
Considering this, we choose a fairly large local context of 30 lines including 18 lines before and 12 lines after the bug location. The slightly asymmetric choice of context window (18 vs. 12) follows previous evidence that found that context before the bug location is slightly more important~\citep{prenner2023out}.

\subsection{Identifier Ingredient Extraction}
The pre-processing step is followed by the ingredient extraction process (i.e., ground-truth ingredients from the ground-truth fix). As mentioned previously, in this work we focus on identifier ingredients. We let \bugIds, \winIds, \methodIds, \fileIds and \projectIds be the set of identifiers occurring in the buggy code lines, the local context (equivalent to a repair model's input window), the surrounding method or function, the file,  the project, respectively and \fixIds the set of identifiers occurring in the fixed version (i.e. the ground-truth fix).
For each bug we extract several sets of identifier ingredients:
\begin{description}
    \item[\fixAll:]  Fix ingredients appear in the fixed lines but not the corresponding buggy code \emph{lines} (\fixIds $\setminus$ \bugIds).
    \item[\winIn:] In-window ingredients appear in the fixed code lines \emph{and} the model's \emph{input window}; e.g., 18 lines before and 12 lines after the buggy line (\fixAll $\cap$ \winIds).
    \item[\winOut:] Out-of-window ingredients appear in the fixed code lines but \emph{not} the \emph{input window} (\fixAll $\setminus$ \winIds).
    \item[\methodIn:] In-method ingredients appear in the fix \emph{and} the \emph{surrounding method or function} (\fixAll $\cap$ \methodIds).
    \item[\methodOut:] Out-of-method ingredients appear in the fix but \emph{not} the \emph{surrounding method or function} (\fixAll $\setminus$ \methodIds). 
    \item[\fileIn:] In-file ingredients appear in the fix \emph{and} the \emph{file context}; i.e., in the file where the bug is located (\fixAll $\cap$ \fileIds).
    \item[\fileOut:] Out-of-file ingredients appear in the fix but \emph{not} the \emph{file context} (\fixAll $\setminus$ \fileIds).
    \item[\projectIn:] In-project ingredients appear in the fix \emph{and} the \emph{project} (\fixAll $\cap$ \projectIds; estimated on a sample for TSSB-3M).
    \item[\projectOut:] Out-of-project ingredients appear in the fix but \emph{not} the \emph{project} (\fixAll $\setminus$ \projectIds; sample estimate for TSSB-3M).    
\end{description}

We extract identifiers (variable names, class names, function and method names, parameter names and field and attribute names) at a \emph{a lexical level} with Pygments\cite{pygments}. Importantly, we do not differentiate between declaration and usage of an identifier, we merely consider their \emph{occurrence}. We also do not discriminate between internal identifiers (i.e., methods or classes declared in project under repair) and external ones (e.g., classes and methods that are part of the standard library or a 3\textsuperscript{rd} party dependencies); we later use manual analysis for that.

\myparagraph{Project-level ingredients.}
To extract project-level ingredients for Defects4J, we rely on Defect4J's \emph{relevant} classes to determine which files need to be searched for ingredients (\texttt{classes.\-relevant.\-src} and \texttt{classes.\-relevant.\-test} bug properties); these are all the classes loaded by the JVM at test execution time.
TSSB-3M is restricted to the file level. We select a random sample of 500 bugs with out-of file project ingredients for which we extract the entire project at the fixed version in order to estimate the prevalence of project-level ingredients in this dataset.

\myparagraph{Ingredient distance.}
For each ingredient in \fileIn we moreover calculate the \emph{distance} from the bug location to the closest occurrence in the same file. We measure distance in number of characters. This is a simple and general choice that is independent of a model's or a lexer's particular tokenization. Further, the distance is \emph{signed}: ingredients that appear before the bug location are negative; those appearing after positive.
In case of multiple occurrences, the distance to the closest one (relative to the bug location) is noted.
Ingredient distance is only calculated for single hunk bugs, that is, bugs with a single bug location and bugs with a single fix ingredient, as otherwise we would have multiple distances per bug.

\myparagraph{Ingredient cover.}
For each of the \enquote{inclusive} ingredient sets (i.e., \winIn, \methodIn, \fileIn and \projectIn) we calculate the \emph{ingredient cover}, which we define as $\frac{|x|}{|fix_{all}|}$, where $x$ is one of the sets. The cover expresses the fraction of fix ingredients that can be found (i.e., are declared or used) at a particular context level. It is only defined for bugs that require at least one fix ingredient (i.e., $|fix_{all}| > 0$). Note that Python allows code to appear outside methods. Thus, in TSSB-3M, roughly 25\% of bugs lack a surrounding function or method. We exclude these bugs when calculating TSSB-3M's method-level cover.

\myparagraph{Uncovered ingredients.} Both datasets contain bugs with uncovered identifiers, who appear in the fixed code but have no occurrences at file (TSSB-3M) or project level (Defects4J).

\myparagraph{Uncovered ingredients in Defects4J.} We use \texttt{ctags} to further analyze these uncovered cases \cite{ctags}. \texttt{ctags} creates an index of all method, class and variable declarations it finds in a file or code base. This allows us to spot \enquote{false positives}: ingredients that appear only in the fixed version, but belong to new variables and methods that are defined inside the code that patches the bug; their scope is usually limited to the patch.

\myparagraph{Uncovered TSSB-3M ingredients.} TSSB-3M does not provide project-level code. Due to its considerable size, we only mine file-level context. To analyze uncovered ingredients, we mine the entire code repository for 500 randomly sampled bugs with \fileOut ingredients (before the fixing commit) and extract project-level identifiers. We then manually analyze cases where \fileOut ingredients are still uncovered at the full project level.

\subsection{Results}\label{sec:res-rq1}

 \myparagraph{Prevalence of Identifier Ingredients.} The number of different types of identifier ingredients differs strongly between TSSB-3M and Defects4J.
In the former, \rqiTssbRatioWithNoIngredient{} of the fixes do not require any identifier ingredient; such bugs include operator fixes (e.g., from {\btHL[fill=red!30]<} to {\btHL[fill=green!30]<=}) or wrapping a statement with a null check. Conversely, in Defects4J only \rqiDfjRatioWithNoIngredient{} of bugs involve no identifier ingredients. The difference is not surprising, since TSSB-3M is a dataset of single statement bugs, while Defects4J features more complex bugs.

\begin{figure}[tbhp]
    \centering
    \includegraphics[width=.9\columnwidth]{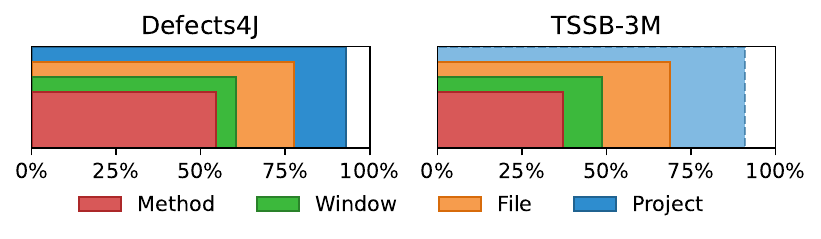}
    \caption{Percentage of fix ingredients covered at the method/function, input window, file, and project level. For TSSB-3M we estimate the project-level on a  sample of 500 bugs with out-of-file ingredients; for Defects4J, we use all \textit{relevant} files belonging to the bug.}
    \label{fig:ingr-cover}
\end{figure}

\myparagraph{Where are the Identifiers?}
We calculate the cover for all bugs in Defects4J and TSSB-3M and visualize it in Figure~\ref{fig:ingr-cover}. We find that the method-context covers \rqiDfjCoverIM{} and \rqiTssbCoverIM{}, the input window's local context \rqiDfjCoverIC{} and \rqiTssbCoverIC{}, the file context \rqiDfjCoverIF{} and \rqiTssbCoverIF{}, and the project context \rqiDfjCoverIT{} and \rqiTssbCoverIT{} of ingredients for Defects4J and TSSB-3M, respectively. Thus, a medium-sized local context of 30 lines contains, on average, slightly more fix-relevant identifiers than the method under repair. On the other hand, 39\% of TSSB-3M identifiers, and 51\% of Defects4J identifiers are not found in a typical input window.

\myparagraph{What Identifiers are Missing?}
%\myparagraph{Defects4J.}
In Defects4J, 7\% of bugs contain uncovered ingredients, that is, ingredients that are not covered even at the project level (white margin in Figure~\ref{fig:ingr-cover}). We find the majority (72\% of bugs) of uncovered ingredients to be false ingredients, that is, identifiers that are newly defined inside the patch's code (e.g. if the patch defines and uses a new variable).
The remaining 28\% of missing ingredients refer to
\begin{inparaenum}[i)]
    \item the standard library (23\%) or
    \item identifiers found in either the project but not the selected subset of relevant project files or identifiers referring to third-party dependencies (<5\%).
\end{inparaenum} 

For TSSB-3M, we examine the 500 bugs with out-of-file ingredients for which we mined full project-level context. In 71\% of cases, the missing identifiers appears at the project level. The remaining 29\% of bugs contained identifiers not found at the project level. Our manual analysis shows that 49\% of these cases contain fix identifiers referring to Python's standard library and 29\% to identifiers of third-party dependencies.  
We see only seven false ingredients (< 5\%, such as iterator variables in newly introduced for loops).
The remaining 17\% of cases with uncovered ingredients can be classified into one of the following cases: \begin{inparaenum}[i)]
\item accesses to attributes which are likely dynamically created (e.g., from database schemas),
\item filenames in import statements,
\item method names never explicitly occurring elsewhere in the project (methods are possibly called by external code or dynamically \enquote{by string}),
\item and test method names.
\end{inparaenum}

\begin{answerbox}{Answer to RQ1}
\emph{Identifier ingredients are prevalent:} even the simple, single statement bugs in TSSB-3M require fix ingredients in almost half of the cases (44\%). This percentage is much higher for the more complex fixes in Defects4J: 85\% of those fixes include identifier ingredients.

\emph{Likely out-of context ingredients are common:} even with a reasonably large context window (18 lines before, 12 after), 39 to 51\% of identifier ingredients are uncovered. These are most often found in the local file, or in the project (over 90\% of fix identifier ingredients appear somewhere in the project).   
Identifiers with no project-wide occurrence mostly belong to standard or third party libraries.
\end{answerbox}

\section{Repair Success (RQ2)}\label{sec:rq2}

In RQ2 we study how identifier ingredients affect repair performance. For Defects4J, we investigate whether the performance of various APR and NPR systems in the recent literature is affected by the number of identifier ingredients or their distance to the fault location. For TSSB-3M, we conduct a similar study while training a new model. We find that the number of fix ingredients, the distance of the ingredients to the bug location (and in particular whether the distance falls within context boundaries) as well as the (absolute) frequency of ingredients in the training set all contribute to repair success.

\subsection{Defects4J}
For Defects4J, we collect data from recent APR and in particular NPR literature and analyze it with respect to identifier ingredient usage and repair success. We analyze repair tools for which repair success on individual bugs was reported, resulting in the following list: TBar~\citep{liuTBarRevisitingTemplatebased2019}, DLFix~\citep{liDLFixContextbasedCode2020}, SequenceR~\citep{chenSequenceRSequencetoSequenceLearning2021}, CIRCLE~\citep{yuanCIRCLEContinualRepair2022}, CoCoNuT~\citep{lutellierCoCoNuTCombiningContextaware2020}, Recoder~\citep{zhuSyntaxguidedEditDecoder2021}, KNOD~\citep{jiangKNODDomainKnowledge2023}, TransplantFix~\citep{yangTransplantFixGraphDifferencingbased2023}, GAMMA\citep{zhangGammaRevisitingTemplateBased2023}, CURE~\citep{jiangCURECodeAwareNeural2021a}, AlphaRepair~\citep{xiaLessTrainingMore2022a}, RAP-Gen~\citep{wangRAPGenRetrievalAugmentedPatch2023a}, FitRepair~\citep{xiaRevisitingPlasticSurgery2023}, RewardRepair~\citep{yeNeuralProgramRepair2022} and TARE~\citep{zhuTareTypeAwareNeural2023}. We then cross this information (i.e., whether a given bug was fixed by a particular NPR tool) with the ingredient-related metrics we computed in RQ1.

\subsection{TSSB-3M}
We use TSSB-3M to \emph{train} specific models: we first fine-tune and then evaluate our own seq2seq repair model on the TSSB-3M-based dataset from RQ1 (e.g., data is deduplicated, see \ref{sec:dedup}). Table~\ref{tab:datasets}) shows the dataset splits we use.

\begin{table}[htbp]
  \footnotesize
  \caption{Dataset splits in this work along with their size and usage. Note that RQ2 and RQ3 use the same evaluation set and that the training sets for RQ3 and RQ4 are disjoint subsets of the training set of RQ2.}
  \label{tab:datasets}
  \begin{tabular}{ccrl}
    \toprule
    Dataset & Language & Size & Use \\
    \midrule
    Defects4J~\citep{justDefects4JDatabaseExisting2014} & Java & 
    832 & Analysis (RQ1, RQ2) \\
    \midrule
    \multirow{7}{*}{TSSM-3M~\citep{richterTSSB3MMiningSingle2022}} & \multirow{7}{*}{Python} & 853,943 & Analysis (RQ1) \\
    & & 510,851 & Training (RQ2) \\
    & & 85,776 & Evaluation (RQ2) \\
    & & 274,776 & Training (Scanner, RQ3) \\
    & & 85,776 & Evaluation (Scanner, RQ3) \\
    & & 236,075 & Training (Repair, RQ4) \\
    & & 257,316 & Evaluation (Repair, RQ4) \\
  \bottomrule
\end{tabular}
\end{table}

\myparagraph{Repair Model}
Our TSSB-3M repair model is based on CodeT5~\citep{wangCodeT5IdentifierawareUnified2021}, a seq2seq foundation model for code (we use the small variant with 60M parameters). The model is fed a buggy code snippet (buggy line of code along with local context) and trained to output fixed code (see below for details on the input format). We use a learning rate of \SI{1e-4} over 4 epochs with a batch size of 12, accumulated over 2 steps. For evaluation, we generate five fix candidates for each bug in the test set using beam search with five beams.

\myparagraph{Model input \& output}\label{sec:rq1-model-input}
The input is the buggy code section, together with 30 lines of local context code (18 before and 12 after the bug location). As is common in recent NPR work~\citep{chenSequenceRSequencetoSequenceLearning2021, prenner2023out, yeNeuralProgramRepair2022, jiangCURECodeAwareNeural2021a} we work under the perfect fault localization assumption and the bug location is marked with special tokens (see Figure~\ref{fig:rq1-input}). The model should output a fixed version of the code portion between the markers. Consequently, the learning target is the code that appears between the markers in the fixed ground-truth snippet.

\begin{figure}[htbp]
\centering
\begin{lstlisting}[language=Python,style=patch,numbers=none,xleftmargin=0pt,xrightmargin=0pt,framesep=0pt,morekeywords={BUGSTART,BUGEND, SCAN}]
(* \textellipsis *)
def cluster_update(self, context, identity, profile_id):
    # Get database representation of the existing cluster
    db_cluster = self.cluster_find(context, identity)
<BUGSTART>
    db_profile = self.profile_fine(context, profile_id)
<BUGEND>
    LOG.info(_LI('Updating cluster %s'), db_cluster.name)
    cluster = cluster_mod.Cluster.load(context, 
                                        cluster=db_cluster)
    (* \textellipsis *)
\end{lstlisting}
\caption{Example input for the repair model. The bug location is marked with special \texttt{<BUGSTART>} and \texttt{<BUGEND>} tokens. Note that here parts of the local contexts are omitted.}
\label{fig:rq1-input}
\end{figure}

\myparagraph{Evaluation}
Since bugs in TSSB-3M are not executable (and lack tests) we resort to \emph{exact match} to assess whether a generated patch is correct or not. We generate 5 fix candidates for each bug, then analyze repair success relative to ingredient-related \enquote{bug properties} (e.g., the number of identifier ingredients needed for a fix or their distance from the bug location).

\subsection{Results: Defects4J}\label{sec:res-rq2}

\begin{figure}[tbhp]
    \centering
    \includegraphics[width=\columnwidth]{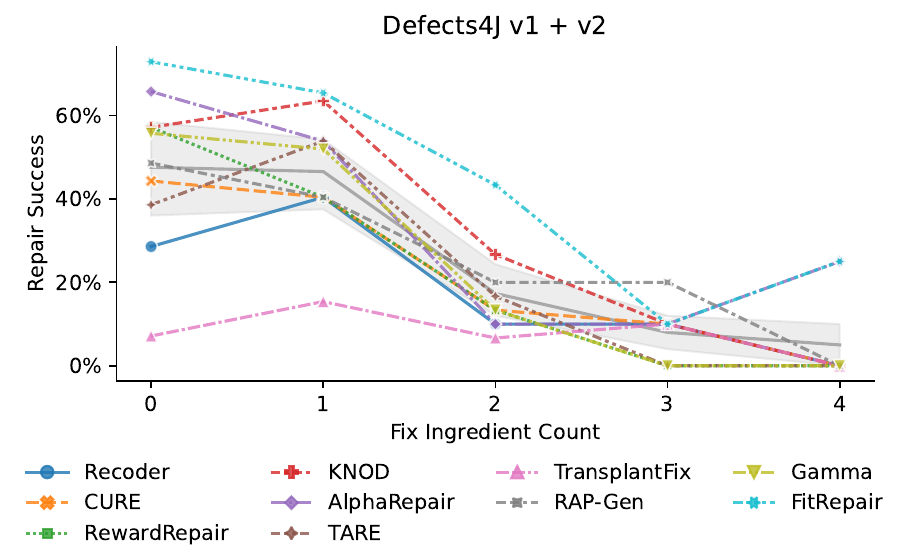}
    \caption{Repair success of APR tools as a function of fix ingredient count for single change bugs in Defects4J. The gray line is the mean over all tools with a 95\% CI band. For most tools, repair success decreases when the number of required fix ingredients increases.
    }
    \label{fig:d4j-perf}
\end{figure}

\myparagraph{Per-tool performance.}
Figure~\ref{fig:d4j-perf} shows repair performance on Defect4J bugs as a function of fix ingredient count for a selection of APR and NPR tools from the literature.
Tools that have been evaluated only on version 1 of Defects4J exhibit a very similar trend but are not depicted for lack of space.
In order to separate the effect of bug complexity (changed lines) and ingredient count, we only use bugs with change count 2, which corresponds to two changed lines in the corresponding fixing patch (e.g., a single replacement, two deletions, two additions). Almost all tools we survey show a downward trend: they seem to have difficulty with bugs that require a high number of fix identifiers. 
TransplantFix~\citep{yangTransplantFixGraphDifferencingbased2023} is an outlier, showing steady (but modest) performance for different fix ingredient counts.

\myparagraph{Performance \& change count.}
While Figure~\ref{fig:d4j-perf} depicts only bugs with a change count of two, we note that similar trends apply for a broader range of change counts: repair success steadily drops as the number of fix ingredients increase.

\subsection{Results on TSSB-3M} 

\begin{figure}[tbhp]
    \centering
    \includegraphics[width=\columnwidth]{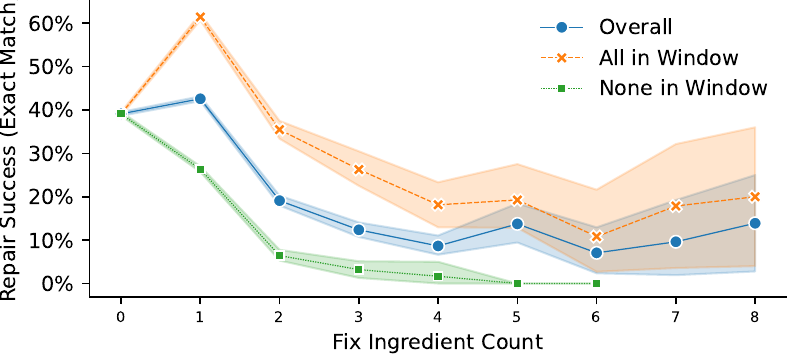}
    \caption{Repair success of bugs in our validation set of TSSB-3M as a function of fix ingredient count and proportion of ingredients that appear within the input window (i.e. the local context). The graph indicates that repair success is highest for single-ingredient bugs, but only for in-context identifiers. Error bands indicate a 95\% confidence interval.
    }
    \label{fig:tssb-perf}
\end{figure}

\myparagraph{Overall performance.} For TSSB-3M we observe a very similar trend: performance decreases as the number of fix ingredients increases (Figure~\ref{fig:tssb-perf}).
Because the bugs in TSSB-3M are all simple single-statement bugs and thus of similar complexity we do not normalize over change count. Performance is clearly higher when all the identifier ingredients are in the context window (orange line), than when they are not (green line).

We notice that performance is rising from bugs with no fix ingredients to single-ingredient bugs (an effect also seen in several tools in Figure~\ref{fig:d4j-perf}). This is strongest with only \winIn ingredients (orange line), and disappears with only \winOut ingredients (green line). This further highlights the strong connection between context and ingredients.

\begin{figure}[tbhp]
    \centering
    \includegraphics[width=\columnwidth]{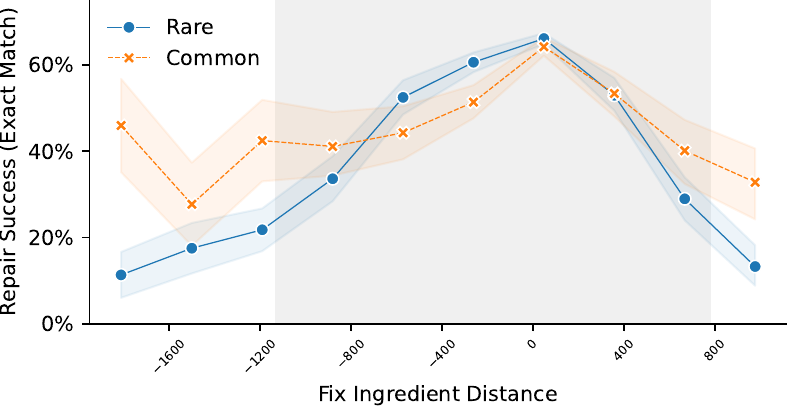}
    \caption{Repair success of bugs with single in-file ingredient in our validation set of TSSB-3M as a function of distance (number of characters) from the bug location and frequency in the training set. Both, frequency and distance to the bug location seems to affect repair success. A distance-related performance drop is most pronounced for bugs with rare fix ingredients. Shaded area shows the 95\textsuperscript{th} percentile of local context size. Error bands indicate a 95\% confidence interval.}
    \label{fig:tssb-perf-dist}
\end{figure}

\myparagraph{Distance \& frequency}
We also analyze repair success in terms of \begin{inparaenum}[i)]
    \item ingredient distance (from the bug location) and
    \item ingredient frequency in the training set.
\end{inparaenum}
Figure~\ref{fig:tssb-perf-dist} summarizes the results. Because aggregating distances of multiple fix ingredients could distort results we only show bugs with a single fix ingredient. In case of multiple occurrences of the fix ingredient, we use the distance \emph{closest} to the bug location. Distance is grouped into 20 equally sized bins; error bands indicate variation inside the bin (95\% CI). Due to outliers, we omit distances below the 10\textsuperscript{th} and above the 90\textsuperscript{th} percentiles. We define an identifier ingredient as \enquote{rare} if it has 50 or less occurrences in the training set, \enquote{common} if it has 500 or more.
The gray shaded area indicates the \emph{median} context window size. More precisely, the area ranges from the median pre-context size (context leading the bug location) to the median post-context size (context following the bug location) in characters.

\myparagraph{Distance affects repair success}
Still looking at Figure~\ref{fig:tssb-perf-dist}, we can see that performance decreases steadily in both directions as the ingredient position moves away from the bug location (0). 
Performance seems to fall off more steeply for ingredients that occurs after the bug location (distance > 0). This may be related to previous work that suggests that context preceding the bug location is slightly more important for repair success than context following it~\citep{prenner2023out}.

\myparagraph{Frequency affects repair success} 
Finally, we observe that the drop in performance is less pronounced for common ingredients. We speculate that for common ingredients the model is less dependent on retrieving ingredients from the context as they might have been (to some degree) \enquote{encoded} into model weights during training.

\begin{answerbox}{Answer to RQ2}
Our findings suggest that fix ingredients have a considerable impact on repair performance. Not only the \emph{number} of fix ingredients affects repair success but also their \emph{distance} to the repair location, their \emph{frequency} of occurrence in the training set and in particular whether they appear \emph{within the local context or not}.
\end{answerbox}

\section{Identifier Extraction Model (RQ3)}\label{sec:rq3}
In RQ3 we explore whether a scanner model (orange in Figure~\ref{fig:scan-fix-overview}) can predict useful ingredients from a wider context that does not fit a repair model's input window (e.g., ingredients in \winOut). We find that our scanner \emph{can} extract fix ingredients from this larger context, albeit with modest performance.

\subsection{Methodology}

\myparagraph{Scanner model.}
To implement this model we rely on an encoder-only (i.e. BERT-like) Transformer. We base our model on BigCode's pre-trained StarEncoder model \cite{li2023starcoder, starEncoder} with roughly 125M parameters. The task of extracting ingredients is framed as a \emph{token classification task}. Tokens (or subtokens) that are (or are part of) a fix identifier are assigned a positive label, all other tokens are assigned a negative label. Assignment of a label happens on a lexical level, so we do not take scope or accessibility into account. We take care to use different training sets for the scanner model and the final ScanFix model (RQ4) to avoid data leakage issues (see Table~\ref{tab:datasets}).

\myparagraph{Model variants.}
We experiment with two variants. The \enquote{All} variant predicts fix ingredients of any type (irrespective of whether they appear in-window or out-of-window). The second variant, \enquote{OOW} predicts only out-of-window ingredients. Both variants are trained in the same way with one difference: for \enquote{All}, \fixAll ingredient tokens have a positive label, while for \enquote{OOW}, only \winOut ingredient tokens have a positive label.

\begin{figure}[htbp]
\centering
\begin{lstlisting}[language=Python,style=patch,numbers=none,xleftmargin=0pt,xrightmargin=0pt,framesep=0pt,morekeywords={BUGSTART,BUGEND, SCAN}]
(* \textellipsis *)
def cluster_update(self, context, identity, profile_id):
    # Get  database representation of the existing cluster
    db_cluster = self.cluster_find(context, identity)
<BUGSTART>
    db_profile = self.profile_fine(context, profile_id)
<BUGEND>
    LOG.info(_LI('Updating cluster %s'), db_cluster.name)
    cluster = cluster_mod.Cluster.load(context, 
                                        cluster=db_cluster)
    (* \textellipsis *)
<SCAN>
(* \textellipsis *)
def profile_type_template(self, context, type_name):
    return {}
def ~~profile_find~~(self, context, identity, show_deleted=False):
    '''Find a profile with the given identity (name or ID).'''
    if uuidutils.is_uuid_like(identity):
        profile = db_api.profile_get(context, identity,
                                show_deleted=show_deleted)
    (* \textellipsis *)                                     
\end{lstlisting}
\caption{Example input for scanner model. The first part is the bug and its local context. The \texttt{<SCAN>} divider token follows, then the code to scan. To fix the bug, \texttt{profile\_find} should be called instead of \texttt{profile\_fin\underline{e}}. The correct identifier ingredient is assigned a positive label in the scan code (orange highlight). All other tokens have negative labels. Portions of local context and scan code omitted for brevity.}
\label{fig:scanner-structure}
\end{figure}

\myparagraph{Model input.}\label{sec:scanner-model-input}
Just like the repair model, the scanner model has a finite input window (1024 tokens). 
The scanner also needs the actual bug as input so it can determine useful repair ingredients relative to this bug. Thus, we divide the input into two parts. The first part contains the buggy code section for which ingredients are sought, together with the local context (as described in Section~\ref{sec:rq1-model-input}).
The second part of the input is a code snippet from which identifiers should be extracted, in other words the code we want to \enquote{scan}; we call it the \emph{scan code}. The scan code snippet fills the remaining space in the input window. The two parts (the bug and the scan code) are separated by a special \texttt{<SCAN>} token. With this input structure, we can scan an arbitrary amount of code by splitting it into any number of appropriately sized snippets. Thus for a single bug sample in our dataset we might end up with several \emph{scanner samples}, each with the same bug prefix (first part) but different scan code (second part). Figure~\ref{fig:scanner-structure} shows such a scanner sample. In practice, since TSSB-3M is limited to a file-level context, we take the scan code from the buggy file.

\myparagraph{Undersampling \& overlapping.} Many samples have no positive label (i.e., the code contains no fix ingredients) leading to class imbalances. We apply undersampling to balance the number of samples with and without positive labels. For the \enquote{OOW} variant, only \winOut ingredients are assigned a positive class. Consequently, there are fewer samples with at least one positive token. To remedy this, we split the scan code differently for each variant. For \enquote{OOW}, we split scan code of a file into overlapping chunks (30\% overlap); for \enquote{All}, we split it into disjoint chunks. This way, we train and evaluate both variants on a similar number of samples.

\myparagraph{Training.}   
We fine-tune our model on scanner samples generated from the bug samples in our training set (see Table~\ref{tab:datasets}). We use a learning rate of \SI{6e-5} and train for 4 epochs with a batch size of 30, accumulated over 3 steps.

\myparagraph{Evaluation.}\label{sec:scanner-evaluation} We evaluate the scanner with precision, recall, and F-score:
\scalebox{0.7}{\parbox{\columnwidth}{%
\begin{gather*}
  Precision = \frac{|\hat{F} \cap F|}{|\hat{F}|}\qquad Recall = \frac{|\hat{F} \cap F|}{|F|} \\
F1 = \frac{1}{\alpha / Precision + (1 - \alpha) / Recall}\quad(\alpha = 0.5)
\end{gather*}
}}

where $\hat{F}$ is the set of extracted identifier ingredients and $F$ is the reference set, that is the set of true fix identifiers. 
We use $F =$ \fixAll for the \enquote{All} variant, and $F =$ \winOut for \enquote{OOW}. For each input token, the scanner outputs a probability score. 
We need to choose a threshold at which we assign a token to the positive class (i.e. a useful ingredient). Thresholds closer to zero will favor recall over precision, conversely, a value close to one trades precision for recall. We experiment with several different thresholds and report results for each of them.
Since for each bug we can have multiple scanner samples (each with a different scan code), we obtain the final set of extracted identifiers ($\hat{F}$ above) for a particular bug by combining the results of each sample through a set-union operation.

\subsection{Results}\label{sec:res-rq3}

\begin{figure}[tbhp]
    \centering
    \includegraphics[width=\columnwidth]{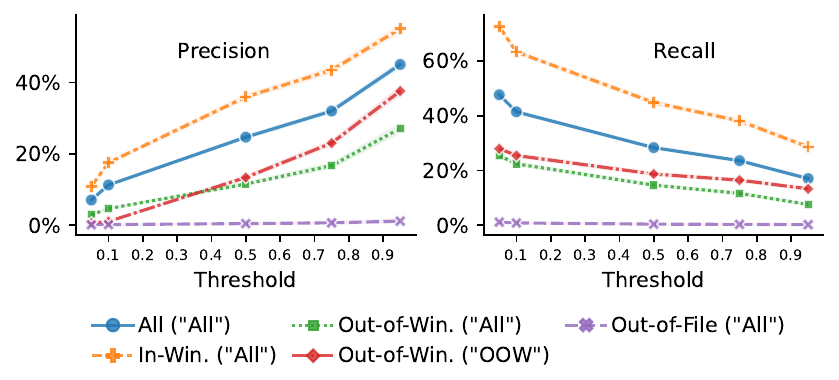}
    \caption{Recall and precision for the TSSB-3M validation set for different thresholds (x-axis), scanner variants (in parentheses) and ingredient sets (curves). Since TSSB-3M is limited to file-level context, the model does not to extract \fileOut ingredients.}
    \label{fig:tssb-scanner-perf}
\end{figure}

\myparagraph{Effect of Threshold.}
The scanner assigns a probability that each token is an ingredient. 
The threshold controls the trade-off between precision and recall. Figure~\ref{fig:tssb-scanner-perf} shows recall and precision for five selected thresholds. As we will see later (RQ4, \ref{tab:rq4-results}), a scanner with a low threshold (high recall, low precision), seems to perform best when combined with a repair model.

\myparagraph{\enquote{All} variant.}
This variant predicts both in-window and out-of-window ingredients. At threshold $t = 0.5$, we obtain a recall of \percentage{0.476631}  and a precision of only \percentage{0.070490}. On the other hand, at a very high threshold of $y = 0.95$, precision rises to \percentage{0.450880}, but recall drops to \percentage{0.170106}. 
When evaluating the model only on out-of-window ingredients, performance is considerably lower: recall and precision are \percentage{0.254100} and \percentage{0.028985} for $t = 0.05$ and, for $t=0.95$, \percentage{0.075414} and \percentage{0.271459}, respectively.

\myparagraph{\enquote{OOW} variant.}
Not surprisingly, our \enquote{OOW} variant is slightly better at predicting \emph{out-of-window ingredients}, with a recall of \percentage{0.279579} at $t=0.05$ and \percentage{0.132687} at $t = 0.95$. Precision ranges from \percentage{0.005925} to \percentage{0.376126} for thresholds $t=0.05$ and $t=0.95$, respectively.

\myparagraph{Are the extracted ingredients useful?}
Recall and precision only give a limited picture of the quality of the extracted ingredients. Next (RQ4, Table~\ref{tab:rq4-results}), we will see that even with modest precision and recall, the scanner can make an important contribution to repair success, especially for \winOut ingredients. 

\begin{answerbox}{Answer to RQ3}
Our StarEncoder-based scanner model is able to extract identifier ingredients with \emph{mixed results}. For the thresholds $t=0.05$ and $t=0.95$, recall ranges from \percentageint{0.476631} to \percentageint{0.170106} and precision from \percentageint{0.070490} to \percentageint{0.450880} for \fixAll ingredients and from \percentageint{0.279579} to \percentageint{0.132687} (recall) and 
\percentageint{0.005925} to \percentageint{0.376126} (precision) for \winOut ingredients. %(with the \enquote{OOC} model variant). 
As later results show, even such modest performance is useful.
\end{answerbox}

\section{ScanFix -- Combined Repair Model (RQ4)}\label{sec:rq4}
In this section, we present evaluation results for our final research question, namely whether we can improve repair success by passing to a repair model identifier ingredients that have been extracted by RQ3's scanner. We call this setup ScanFix. 

\subsection{Methodology}\label{sec:meth-rq4}
\myparagraph{ScanFix.}
As indicated in Figure~\ref{fig:scan-fix-overview}, we combine the scanner model from RQ3 with the repair model used in RQ2. In this setup, the repair process is divided into two steps. In the first step we extract a set of possible fix ingredients by passing a bug with its local context and scan code to the scanner model.
The set of \emph{predicted} fix ingredients is then passed, together with the bug and the local context to the repair model.

\myparagraph{Model input \& training.}
The input format closely follows the one used in RQ2 (see Figure~\ref{fig:rq1-input} and Figure~\ref{fig:repair-with-ingr}) with one important difference. At the beginning of the input window we place the predicted identifier ingredients, separated by space and followed by another separator token (\texttt{<INGR>}). As mentioned in Section~\ref{sec:scanner-evaluation} a threshold $t$ is needed for ingredient extraction. For each of the thresholds $t=0.05$, $t=0.5$, $t=0.95$ we train and evaluate as described in RQ2, albeit with a different training set (i.e., half of the training data is used for the scanner model, the other half for the repair model; see Table~\ref{tab:datasets}). Note that predicted ingredients are placed in the input during training \emph{and} prediction. All training and evaluation steps are carried out for both scanner model variants separately.

\myparagraph{Baselines.}
To compare the performance of ScanFix (i.e., the combined scanner and repair models) we compare their performance with the following baselines. 
\begin{description}
    \item[Perfect ingredients:] a version of ScanFix with a \enquote{perfect} or oracle scanner module. That is, we use the ground-truth fix ingredients instead of the predicted ingredients.
    \item[Perfect ingredients (file):] as above, but limited to \fileIn ingredients. This is the theoretical limit for this version of ScanFix as we scan for ingredients at the file level.
    \item[Perfect recall, low precision:] perfect \fileIn ingredients, but mixed with unrelated identifiers. For each true fix ingredient we add 20 \enquote{useless} ingredients randomly sampled from the file under repair. In terms of ingredients we have perfect recall, but a very low precision of only 5\%.
    \item[Naive ingredients:] A simple selection heuristic. If there is any free space left in the input window after placing the bug location and the local context we fill it up with identifiers from the file-under-repair. Identifiers are not selected by any particular order and only placed at most once.
    \item[No ingredients:] the repair model without a scanner (similar to RQ2's repair model, with a smaller training set).
    \item[Large context:] a repair model without scanner but a very large context. For this model we use 5120 tokens (fully exhausting our VRAM budget) to expand the local context. All other models use an input window of 1024 tokens.
\end{description}

\subsection{Results}

Table~\ref{tab:rq4-results} gives an overview of the results and compares various ScanFix variants with the baselines. Evaluation was done on the TSSB-3M dataset (see Table~\ref{tab:datasets}). 

\myparagraph{Scanned ingredients help.}
We find a significant increase in repair success (exact match), in particular for out-of-window ingredients. When comparing results with a no-augmentation baseline (\enquote{No ingrs.} in Table~\ref{tab:rq4-results}) we see an absolute performance increase of 2.55\% and a relative improvement of roughly 7\%
obtained with our \enquote{All} scanner model at threshold $t=0.05$.
The performance increase is more pronounced for out-of-window ingredients. Here, a repair model augmented with ingredients from the \enquote{OOW} scanner variant can obtain a relative improvement of 31.5\% (abs. 5.9\%).

\myparagraph{Large-context baseline very strong.}
With a relative improvement of 47.8\%, our large-context baseline shows surprisingly strong performance surpassing all of the ScanFix variants.

\myparagraph{Perfect ingredients at the top.}
By a large margin, augmentation with perfect ingredients outperforms all other baselines. 
The baseline with perfect file-level ingredient is the absolute theoretical limit of a scanner model.

\begin{figure}[tbhp]
    \centering
    \includegraphics[width=\columnwidth]{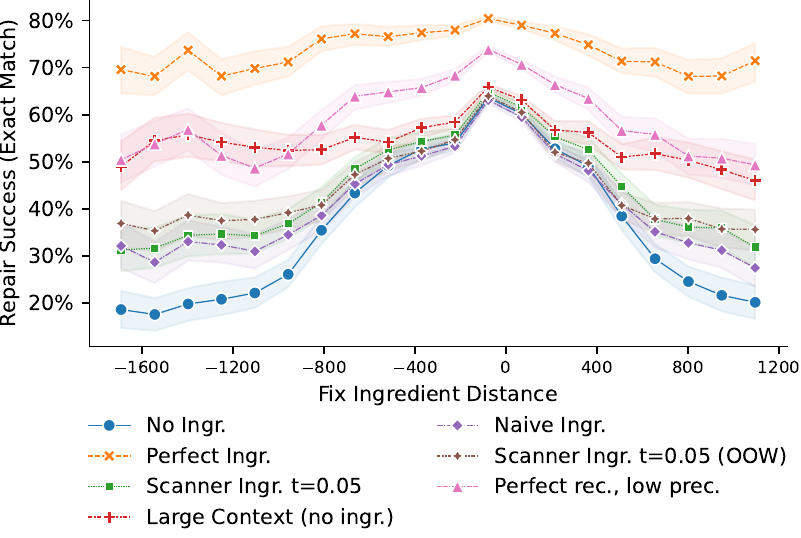}
    \caption{Repair success of bugs with single in-file ingredient in our test set of TSSB-3M as a function of distance (number of characters) from the bug location. Perfect ingredients perform best; the large context model performs better than the best ScanFix variant (t=0.05). Error bands indicate a 95\% CI.}
    \label{fig:tssb-repair-dist}
\end{figure}

\myparagraph{Improvements mostly for far-away ingredients.}
Figure~\ref{fig:tssb-repair-dist} shows how performance varies for different ingredient distances. Apart from the perfect ingredient baseline at the top, we can see that performance differences mostly occur for far-away ingredients which is somewhat expected, as ingredients close to the bug location will be in-window for all models and baseline.

\myparagraph{High precision is not necessary.}
Our baseline with all file-level fix ingredients mixed with additional unrelated ingredients (i.e.,  100\% recall, 5\% precision) fares surprisingly well and outperforms the large context baseline. This shows that the repair model is relatively good at picking the correct ingredients from a large list of choices.

\myparagraph{Low performance for multiple fix ingredients.}
% Figure was cut, rephrase
Figures~\ref{fig:d4j-perf} and \ref{fig:tssb-perf} show that APR tools struggle when ingredient counts increase; ScanFix is no exception (Figure omitted due to space constraints). 
Interestingly, even with perfect ingredients repair performance drops drastically. We conjecture that for bugs with a large number of ingredients the complexity of \enquote{arranging} the ingredients dominates the problem of missing ingredient knowledge. Still, some examples are successful (see Figure~\ref{fig:repair-with-ingr}).

\begin{table}
    \centering
    \footnotesize
    \caption{Repair success for different repair models on our TSSB-3M test set for: all bugs; bugs with at least one ingredient; and bugs with at least one \winOut. Best ScanFix and best non-theoretical baseline results underlined.}
    \label{tab:rq4-results}
    \begin{tabular}{lccc}
        \toprule
        \thead{\large{Model}} & \thead{All\\Bugs} & \thead{Bugs with\\ \fixAll Ingr.} & \thead{Bugs with\\ \winOut Ingr.} \\
        \midrule
       Perfect Ingrs.  & \percentage{0.519855} & \percentage{0.692558} & \percentage{0.652346} \\
       Perfect Ingrs. (file) & \percentage{0.442460} & \percentage{0.531326} & \percentage{0.362571} \\
       Perfect Recall, low Precision & \percentage{0.402936} & \percentage{0.447029} & \percentage{0.288114} \\       
       \midrule
       Scanner Ingrs. t=0.05 & \underline{\percentage{0.372728}} & \percentage{0.374321} & \percentage{0.232280} \\
       Scanner Ingrs. t=0.05 (OOW) & \percentage{0.366716} & \underline{\percentage{0.375078}} & \underline{\percentage{0.245616}} \\       
       Scanner Ingrs. t=0.5 & \percentage{0.364921} & \percentage{0.367056} & \percentage{0.222043} \\
       Scanner Ingrs. t=0.5 (OOW) & \percentage{0.362189} & \percentage{0.369240} & \percentage{0.234900} \\       
       Scanner Ingrs. t=0.95 & \percentage{0.356764} & \percentage{0.356630} & \percentage{0.207797} \\
       Scanner Ingrs. t=0.95 (OOW) & \percentage{0.359189} & \percentage{0.364864} & \percentage{0.223880} \\  
       \midrule
       Large Context (no ingrs.) & \underline{\percentage{0.392789}} & \underline{\percentage{0.404331}} & \underline{\percentage{0.275961}} \\
       Naive Ingrs. & \percentage{0.348731}  & \percentage{0.351505} & \percentage{0.213738} \\
       No Ingrs.  &  \percentage{0.347211} & \percentage{0.341898} & \percentage{0.186811} \\
       \bottomrule
    \end{tabular}
\end{table}

\begin{figure}[htbp]
\centering
\begin{subfigure}{\columnwidth}
\begin{lstlisting}[language=Python,style=patch,numbers=none,xleftmargin=0pt,xrightmargin=0pt,framesep=0pt,morekeywords={BUGSTART,BUGEND, INGRE}]
to_str six text_type specs tostring ElementTree xml out 
cpu_specs findall cpu toxml find model text get cpu_model
full str xmlutil <INGRE>
      (*\textellipsis*)
      if model_node is None:
        cpu_model = None
      else:
        cpu_model = model_node.get("name")
      cpu.extend([feature \
            for feature in cpu_specs.findall("feature")])
  if out == "salt":
      return {
        "model": cpu.find("model").text,
        "vendor": cpu.find("vendor").text,
        "features": [feature.get("name") \
            for feature in cpu.findall("feature")],
      }
<BUGSTART>
  @return cpu.toxml()@
<BUGEND>
def network_define(* \textellipsis *)
\end{lstlisting}
\caption{Model input}
\end{subfigure}

\begin{subfigure}{\columnwidth}
\begin{lstlisting}[language=Python,style=patch,numbers=none,xleftmargin=0pt,xrightmargin=0pt,framesep=0pt,morekeywords={BUGSTART,BUGEND, INGRE}]
(*\xmarkrectt{1}*)return xmlutil.tostring(cpu.toxml())
(*\xmarkrectt{2}*)return xmlutil.tostring(cpu)
(*\xmarkrectt{3}*)return xml.tostring(cpu)
(*\cmarkrectt{4}*)return ElementTree.tostring(cpu)
(*\xmarkrectt{5}*)return xmlutil.toxml(cpu)
\end{lstlisting}
\caption{Model predictions}
\end{subfigure}
% BUG LOC 3005
\caption{Example of a bug in TSSB-3M that was correctly fixed with the help of the scanner module. The correct prediction (\#4) requires two out-of-window ingredients (\texttt{ElementTree} and \texttt{tostring}), both proposed by the scanner (before \texttt{<INGRE>} token). Parts of the input have been omitted and/or reformatted.}
\label{fig:repair-with-ingr}
\end{figure}

\begin{answerbox}{Answer to RQ4}
ScanFix (ingredients from scanner model + repair model) performs  better than the baselines without any ingredients as well as baselines with randomly selected ingredients, especially for bugs with out-of-window fix ingredients. However, ScanFix' performance remains behind the large context baseline.
\end{answerbox}

\section{Discussion}\label{sec:implications}

\subsection{Limitations}
\label{sec:limitations}
\myparagraph{Bugs.} Despite due diligence we cannot fully preclude software bugs in our training and evaluation scripts. 

\myparagraph{Non-Identifier Ingredients.}
This work is limited to identifier ingredients. Literals (e.g. integer constants, specific bitmasks, strings, dictionary keys) as well as compound ingredients, such as whole statements or short idiomatic code snippets are other forms of important ingredients that are not studied in this work.   

\myparagraph{Lexical Analysis.}
In order to determine identifier ingredients we resort to lexical analysis. This type of analysis is fairly crude in that it ignores for instance an identifier's syntactic function (e.g., a variable name vs. a class or method name) as well as its scope (e.g., is a certain variable even accessible). Incorporating such information by either filtering extracted ingredients (e.g., discarding inaccessible or incompatible identifiers) or direct integration into the training process could improve performance. This, however, would be connected with a substantial engineering effort beyond the scope of this work.

\myparagraph{Single model.} We use a single model architecture for our scanner as well as for our repair models. Our repair model is based on T5/CodeT5. This architecture was used in a wide range of existing work~\citep{berabi2021tfix, prenner2023out, yeSelfAPRSelfsupervisedProgram2023, yeNeuralProgramRepair2022, xiaRevisitingPlasticSurgery2023, zhangCoditT5PretrainingSource2023}. Also, due to the combination of scanner and repair model and their variants, repeating experiments using a second or even third model architecture would have exceeded our computational budget (we must train and evaluate a repair model for each scanner variant and threshold). In particular, for our large context baseline we specifically did not use a different model with larger context (e.g., an LLM such as CodeLlama~\citep{roziereCodeLlamaOpen2024}) but adapted the model type used for the other experiments with an expanded input window to allow an apples-to-apples comparison.

\myparagraph{Impact of LLMs.} Beyond the apples-to-apples comparison, we decided not to focus on LLMs in this work for another reason. Previous work expressed concerns of memorization issues in APR \cite{prennerCanOpenAICodex2022a}, and in SE in general (\cite{al2024traces},\cite{sallou2024breaking}, \cite{karmakar2022codex}), which may lead to over-optimistic performance. Our model is based on the small version of CodeT5 (60M parameters), both due to our limited resources and to minimize these memorization issues. Future work should investigate the impact of identifier ingredients for LLM specifically, while considering memorization issues.

\myparagraph{Datasets.} We chose two dataset to increase the reliability of our findings. However, we were not able to use Defects4J, one of the standard datasets for APR, in RQ3 and RQ4. This is due to its small size. Defects4J has only around 800 bugs. These bugs span all the categories of bugs (e.g., no ingredients, \methodIn ingredients, \fileOut ingredients, etc). This means that looking at subsets of the data, as we do in RQ3 and RQ4, would be very unreliable, as each individual bug would have a very large weight. In contrast, our evaluation set for TSSB is between 80 times (RQ3) and 250 times larger (RQ4), which ensures much more robust results.

\myparagraph{File-level scanning for TSSB-3M.} While TSSB-3M's size allows us to have robust results, one downside is that the sheer number of bugs makes mining for project-level identifiers impractical, as this has to be done for each project versions (thousands of projects, hundreds of thousands of versions). This is why we limit ourselves to \fileIn ingredients. On the other hand, we can not evaluate the performance of the scanner and repair models on \projectIn ingredients. 
As the theoretical baselines show, these are cases where the ScanFix approach has potential; furthermore, it is not obvious how to feed a context larger than a file to a model, beyond RAG approaches \cite{wangRAPGenRetrievalAugmentedPatch2023a} or very large windows.

\subsection{Implications}
\myparagraph{A case of Sutton's bitter lesson?} Sutton's bitter lesson~\citep{sutton2019bitter} states that in the field of machine learning, at least in the long run, computation power trumps domain knowledge. We believe that we have here such a case. Our ScanFix model is outperformed by a large-context baseline. But while the former required a substantial engineering effort, for the latter we only had to increase the input window size (a single configuration option) and \enquote{throw} more computation at the problem. This is a concerning finding that discourages further research into domain-specific solutions for the problem of missing identifiers. 

\myparagraph{Large-context models might solve the ingredient problem.} 
In recent months several new large language models compete on their input window sizes. Code Llama~\citep{roziereCodeLlamaOpen2024} was trained with a context size 16k (and can theoretically handle up to 100k input tokens), while Mistral's Mixtral~\citep{jiangMixtralExperts2024} has a context size of 32k.
Anthropic's Claude family of models (in particular Claude 2 and Claude 3) support context windows of up to 200k tokens~\citep{anthropicClaude}. Finally, Google's recent Gemini 1.5 model boasts millions of input tokens~\citep{reidGeminiUnlockingMultimodal2024}. Input windows of such size allow it to include entire code repositories. 
Given our results, it is possible that large input windows could solve the problem of missing ingredients, making approaches like ScanFix obsolete. 

\myparagraph{Large context might \emph{not} solve the ingredient problem.} Being able to fit an entire repository (or a significant part of it) in a LLM's context window is one thing; using such a large context window effectively is another. Until recently, Transformer-based LLMs made much better use of information that appeared at either the beginning or the end of the input window~\citep{liuLostMiddleHow2023}, for retrieval and question answering tasks. While this \enquote{Lost in the middle} effect is improving for recall tasks (memorizing a specific value in the input window), whether these improvements translate to harder tasks such as identifying which identifier to use is an open question. More complex context benchmarks point to the contrary \cite{hsieh2024ruler}. In addition, one should keep in mind the high computational cost of LLM inference on a large context window, as self-attention is quadratic \cite{vaswaniAttentionAllYou2017}. Sub-quadratic architectures such as Mamba \cite{gu2023mamba} tend to perform worse on these context benchmarks \cite{hsieh2024ruler}.

\myparagraph{Ingredients are still important.} Despite the above discussion, the \enquote{perfect ingredients} baseline shows that even models with limited input window can successfully fix bugs, given that they are provided with the correct identifiers. Moreover, the repair model can handle low precision rather well and is able to select useful ingredients from a large list of ingredient proposals. Whether ingredient extraction models can be substantially improved, and above all, whether in the face of Sutton's bitter lesson, this is a subject worth pursuing, remains an open question. We have identified potential improvements: for instance, we are currently aggregating the results of multiple runs of the scanner via a set union operation; ranking the ingredients via a voting mechanism could lead to better results. In addition, we have not yet looked into whether hybrid approaches that combine ingredient scanning with larger context windows could improve performance, while resulting in smaller overall context windows.

\section{Conclusion}
\label{sec:conclusions}
This work focused on the importance of identifier ingredients (e.g., variables, function, or class names) in NPR, from multiple angles. We first undertook empirical studies on two datasets (TSSB-3M and Defects4J), showing that identifier ingredients are prevalent, and out of context ingredients are common (RQ1). We then showed that identifier ingredients have a considerable impact on performance (RQ2). Based on these insights, we proposed an NPR approach that relies on a scanner module to identify likely ingredients, albeit with modest performance (RQ3). Coupling this scanner to a repair model yielded appreciable performance improvement (RQ4). However, we also found that simply increasing the input size and thus the number of context code lines leads to even larger performance increases (albeit below the theoretical performance of ScanFix with a perfect scanner). As researchers are working to make larger input windows possible and efficient, whether a separate ingredient extraction step as we proposed can establish itself as part of next-generation NPR systems is an open question.
\paragraph*{Acknowledgments}
This study has received financial support from the French State in the framework of the Investments for the Future programme IdEx université de Bordeaux.

\bibliographystyle{IEEETran}
\bibliography{main}

\begin{thebibliography}{10}
\providecommand{\url}[1]{#1}
\csname url@samestyle\endcsname
\providecommand{\newblock}{\relax}
\providecommand{\bibinfo}[2]{#2}
\providecommand{\BIBentrySTDinterwordspacing}{\spaceskip=0pt\relax}
\providecommand{\BIBentryALTinterwordstretchfactor}{4}
\providecommand{\BIBentryALTinterwordspacing}{\spaceskip=\fontdimen2\font plus
\BIBentryALTinterwordstretchfactor\fontdimen3\font minus
  \fontdimen4\font\relax}
\providecommand{\BIBforeignlanguage}[2]{{%
\expandafter\ifx\csname l@#1\endcsname\relax
\typeout{** WARNING: IEEEtran.bst: No hyphenation pattern has been}%
\typeout{** loaded for the language `#1'. Using the pattern for}%
\typeout{** the default language instead.}%
\else
\language=\csname l@#1\endcsname
\fi
#2}}
\providecommand{\BIBdecl}{\relax}
\BIBdecl

\bibitem{lutellierCoCoNuTCombiningContextaware2020}
T.~Lutellier, H.~V. Pham, L.~Pang, Y.~Li, M.~Wei, and L.~Tan, ``{{CoCoNuT}}:
  Combining context-aware neural translation models using ensemble for program
  repair,'' in \emph{Proceedings of the 29th {{ACM SIGSOFT International
  Symposium}} on {{Software Testing}} and {{Analysis}}}, ser. {{ISSTA}}
  2020.\hskip 1em plus 0.5em minus 0.4em\relax ACM Press, Jul. 2020, pp.
  101--114.

\bibitem{prennerCanOpenAICodex2022a}
J.~A. Prenner, H.~Babii, and R.~Robbes, ``Can {{OpenAI}}'s codex fix bugs? an
  evaluation on {{QuixBugs}},'' in \emph{Proceedings of the {{Third
  International Workshop}} on {{Automated Program Repair}}}, ser. {{APR}}
  '22.\hskip 1em plus 0.5em minus 0.4em\relax ACM Press, Oct. 2022, pp. 69--75.

\bibitem{jiangImpactCodeLanguage2023}
N.~Jiang, K.~Liu, T.~Lutellier, and L.~Tan, ``Impact of {{Code Language
  Models}} on {{Automated Program Repair}},'' Apr. 2023.

\bibitem{barrPlasticSurgeryHypothesis2014}
E.~T. Barr, Y.~Brun, P.~Devanbu, M.~Harman, and F.~Sarro, ``The plastic surgery
  hypothesis,'' in \emph{Proceedings of the 22nd {{ACM SIGSOFT International
  Symposium}} on {{Foundations}} of {{Software Engineering}}}, ser. {{FSE}}
  2014.\hskip 1em plus 0.5em minus 0.4em\relax ACM Press, Nov. 2014, pp.
  306--317.

\bibitem{prenner2023out}
J.~A. Prenner and R.~Robbes, ``Out of context: How important is local context
  in neural program repair?'' \emph{arXiv preprint arXiv:2312.04986}, 2023.

\bibitem{justDefects4JDatabaseExisting2014}
R.~Just, D.~Jalali, and M.~D. Ernst, ``{{Defects4J}}: A database of existing
  faults to enable controlled testing studies for {{Java}} programs,'' in
  \emph{Proceedings of the 2014 {{International Symposium}} on {{Software
  Testing}} and {{Analysis}}}, ser. {{ISSTA}} 2014.\hskip 1em plus 0.5em minus
  0.4em\relax ACM Press, Jul. 2014, pp. 437--440.

\bibitem{jFreeChart}
``Jfreechart charting library documentation,''
  https://www.jfree.org/jfreechart/, accessed: 2024-07-19.

\bibitem{chenSequenceRSequencetoSequenceLearning2021}
Z.~Chen, S.~Kommrusch, M.~Tufano, L.-N. Pouchet, D.~Poshyvanyk, and
  M.~Monperrus, ``{{SequenceR}}: {{Sequence-to-Sequence Learning}} for
  {{End-to-End Program Repair}},'' \emph{IEEE Transactions on Software
  Engineering}, vol.~47, no.~9, pp. 1943--1959, Sep. 2021.

\bibitem{yeNeuralProgramRepair2022}
H.~Ye, M.~Martinez, and M.~Monperrus, ``Neural program repair with
  execution-based backpropagation,'' in \emph{Proceedings of the 44th
  {{International Conference}} on {{Software Engineering}}}, ser. {{ICSE}}
  '22.\hskip 1em plus 0.5em minus 0.4em\relax New York, NY, USA: Association
  for Computing Machinery, May 2022, pp. 1506--1518.

\bibitem{xiaRevisitingPlasticSurgery2023}
C.~S. Xia, Y.~Ding, and L.~Zhang, ``Revisiting the {{Plastic Surgery
  Hypothesis}} via {{Large Language Models}},'' Mar. 2023.

\bibitem{richterTSSB3MMiningSingle2022}
C.~Richter and H.~Wehrheim, ``{{TSSB-3M}}: {{Mining}} single statement bugs at
  massive scale,'' in \emph{2022 {{IEEE}}/{{ACM}} 19th {{International
  Conference}} on {{Mining Software Repositories}} ({{MSR}})}, May 2022, pp.
  418--422.

\bibitem{scanfixReplication}
``Replication package for "extracting fix ingredients using language models",''
  https://github.com/giganticode/llm\_ingredient\_extraction.

\bibitem{martinezFixIngredientsAlready2014}
M.~Martinez, W.~Weimer, and M.~Monperrus, ``Do the fix ingredients already
  exist? an empirical inquiry into the redundancy assumptions of program repair
  approaches,'' in \emph{Companion {{Proceedings}} of the 36th {{International
  Conference}} on {{Software Engineering}}}, ser. {{ICSE Companion}}
  2014.\hskip 1em plus 0.5em minus 0.4em\relax ACM Press, May 2014, pp.
  492--495.

\bibitem{gouesGenProgGenericMethod2012}
C.~L. Goues, T.~Nguyen, S.~Forrest, and W.~Weimer, ``{{GenProg}}: {{A Generic
  Method}} for {{Automatic Software Repair}},'' \emph{IEEE Transactions on
  Software Engineering}, vol.~38, no.~1, pp. 54--72, Jan. 2012.

\bibitem{yuanARJAAutomatedRepair2020}
Y.~Yuan and W.~Banzhaf, ``{{ARJA}}: {{Automated Repair}} of {{Java Programs}}
  via {{Multi-Objective Genetic Programming}},'' \emph{IEEE Transactions on
  Software Engineering}, vol.~46, no.~10, pp. 1040--1067, Oct. 2020.

\bibitem{martinezUltraLargeRepairSearch2018}
M.~Martinez and M.~Monperrus, ``Ultra-{{Large Repair Search Space}} with
  {{Automatically Mined Templates}}: The {{Cardumen Mode}} of {{Astor}},''
  \emph{arXiv:1712.03854 [cs]}, vol. 11036, pp. 65--86, 2018.

\bibitem{lewis2020retrieval}
P.~Lewis, E.~Perez, A.~Piktus, F.~Petroni, V.~Karpukhin, N.~Goyal,
  H.~K{\"u}ttler, M.~Lewis, W.-t. Yih, T.~Rockt{\"a}schel \emph{et~al.},
  ``Retrieval-augmented generation for knowledge-intensive nlp tasks,''
  \emph{Advances in Neural Information Processing Systems}, vol.~33, pp.
  9459--9474, 2020.

\bibitem{wangRAPGenRetrievalAugmentedPatch2023a}
W.~Wang, Y.~Wang, S.~Joty, and S.~C. Hoi, ``{{RAP-Gen}}: {{Retrieval-Augmented
  Patch Generation}} with {{CodeT5}} for {{Automatic Program Repair}},'' in
  \emph{Proceedings of the 31st {{ACM Joint European Software Engineering
  Conference}} and {{Symposium}} on the {{Foundations}} of {{Software
  Engineering}}}, ser. {{ESEC}}/{{FSE}} 2023.\hskip 1em plus 0.5em minus
  0.4em\relax ACM Press, Nov. 2023, pp. 146--158.

\bibitem{liuAutomatedCodeEditing2024}
C.~Liu, P.~Cetin, Y.~Patodia, B.~Ray, S.~Chakraborty, and Y.~Ding, ``Automated
  {{Code Editing}} with {{Search-Generate-Modify}},'' in \emph{Proceedings of
  the 2024 {{IEEE}}/{{ACM}} 46th {{International Conference}} on {{Software
  Engineering}}: {{Companion Proceedings}}}, ser. {{ICSE-Companion}} '24.\hskip
  1em plus 0.5em minus 0.4em\relax New York, NY, USA: Association for Computing
  Machinery, May 2024, pp. 398--399.

\bibitem{zhangAutoCodeRoverAutonomousProgram2024}
Y.~Zhang, H.~Ruan, Z.~Fan, and A.~Roychoudhury, ``{{AutoCodeRover}}:
  {{Autonomous Program Improvement}},'' in \emph{Proceedings of the 33rd {{ACM
  SIGSOFT International Symposium}} on {{Software Testing}} and {{Analysis}}},
  ser. {{ISSTA}} 2024.\hskip 1em plus 0.5em minus 0.4em\relax New York, NY,
  USA: Association for Computing Machinery, Sep. 2024, pp. 1592--1604.

\bibitem{bouzeniaRepairAgentAutonomousLLMBased2024}
I.~Bouzenia, P.~Devanbu, and M.~Pradel, ``{{RepairAgent}}: {{An Autonomous}},
  {{LLM-Based Agent}} for {{Program Repair}},'' Mar. 2024.

\bibitem{yeSelfAPRSelfsupervisedProgram2023}
H.~Ye, M.~Martinez, X.~Luo, T.~Zhang, and M.~Monperrus, ``{{SelfAPR}}:
  {{Self-supervised Program Repair}} with {{Test Execution Diagnostics}},'' in
  \emph{Proceedings of the 37th {{IEEE}}/{{ACM International Conference}} on
  {{Automated Software Engineering}}}, ser. {{ASE}} '22.\hskip 1em plus 0.5em
  minus 0.4em\relax ACM Press, Jan. 2023, pp. 1--13.

\bibitem{yangWhereWereRepair2021}
D.~Yang, K.~Liu, D.~Kim, A.~Koyuncu, K.~Kim, H.~Tian, Y.~Lei, X.~Mao, J.~Klein,
  and T.~F. Bissyand{\'e}, ``Where were the repair ingredients for
  {{Defects4j}} bugs?'' \emph{Empirical Software Engineering}, vol.~26, no.~6,
  p. 122, Sep. 2021.

\bibitem{treesitter}
``Tree-sitter parsing library documentation,'' https://tree-sitter.github.io,
  accessed: 2024-07-19.

\bibitem{pygments}
``Pygments lexer library documentation,'' https://pygments.org/, accessed:
  2024-07-19.

\bibitem{ctags}
``Ctags utility github repository,'' https://github.com/universal-ctags/ctags,
  accessed: 2024-07-19.

\bibitem{liuTBarRevisitingTemplatebased2019}
K.~Liu, A.~Koyuncu, D.~Kim, and T.~F. Bissyand{\'e}, ``{{TBar}}: Revisiting
  template-based automated program repair,'' in \emph{Proceedings of the 28th
  {{ACM SIGSOFT International Symposium}} on {{Software Testing}} and
  {{Analysis}}}, ser. {{ISSTA}} 2019.\hskip 1em plus 0.5em minus 0.4em\relax
  ACM Press, Jul. 2019, pp. 31--42.

\bibitem{liDLFixContextbasedCode2020}
Y.~Li, S.~Wang, and T.~N. Nguyen, ``{{DLFix}}: Context-based code
  transformation learning for automated program repair,'' in \emph{Proceedings
  of the {{ACM}}/{{IEEE}} 42nd {{International Conference}} on {{Software
  Engineering}}}, ser. {{ICSE}} '20.\hskip 1em plus 0.5em minus 0.4em\relax ACM
  Press, Jun. 2020, pp. 602--614.

\bibitem{yuanCIRCLEContinualRepair2022}
W.~Yuan, Q.~Zhang, T.~He, C.~Fang, N.~Q.~V. Hung, X.~Hao, and H.~Yin,
  ``{{CIRCLE}}: Continual repair across programming languages,'' in
  \emph{Proceedings of the 31st {{ACM SIGSOFT International Symposium}} on
  {{Software Testing}} and {{Analysis}}}, ser. {{ISSTA}} 2022.\hskip 1em plus
  0.5em minus 0.4em\relax ACM Press, Jul. 2022, pp. 678--690.

\bibitem{zhuSyntaxguidedEditDecoder2021}
Q.~Zhu, Z.~Sun, Y.-a. Xiao, W.~Zhang, K.~Yuan, Y.~Xiong, and L.~Zhang, ``A
  syntax-guided edit decoder for neural program repair,'' in \emph{Proceedings
  of the 29th {{ACM Joint Meeting}} on {{European Software Engineering
  Conference}} and {{Symposium}} on the {{Foundations}} of {{Software
  Engineering}}}, ser. {{ESEC}}/{{FSE}} 2021.\hskip 1em plus 0.5em minus
  0.4em\relax ACM Press, Aug. 2021, pp. 341--353.

\bibitem{jiangKNODDomainKnowledge2023}
N.~Jiang, T.~Lutellier, Y.~Lou, L.~Tan, D.~Goldwasser, and X.~Zhang,
  ``{{KNOD}}: {{Domain Knowledge Distilled Tree Decoder}} for {{Automated
  Program Repair}},'' Apr. 2023.

\bibitem{yangTransplantFixGraphDifferencingbased2023}
D.~Yang, X.~Mao, L.~Chen, X.~Xu, Y.~Lei, D.~Lo, and J.~He, ``{{TransplantFix}}:
  {{Graph Differencing-based Code Transplantation}} for {{Automated Program
  Repair}},'' in \emph{Proceedings of the 37th {{IEEE}}/{{ACM International
  Conference}} on {{Automated Software Engineering}}}, ser. {{ASE}} '22.\hskip
  1em plus 0.5em minus 0.4em\relax ACM Press, Jan. 2023, pp. 1--13.

\bibitem{zhangGammaRevisitingTemplateBased2023}
Q.~Zhang, C.~Fang, T.~Zhang, B.~Yu, W.~Sun, and Z.~Chen, ``Gamma: {{Revisiting
  Template-Based Automated Program Repair Via Mask Prediction}},'' in
  \emph{2023 38th {{IEEE}}/{{ACM International Conference}} on {{Automated
  Software Engineering}} ({{ASE}})}, Sep. 2023, pp. 535--547.

\bibitem{jiangCURECodeAwareNeural2021a}
N.~Jiang, T.~Lutellier, and L.~Tan, ``{{CURE}}: {{Code-Aware Neural Machine
  Translation}} for {{Automatic Program Repair}},'' in \emph{2021
  {{IEEE}}/{{ACM}} 43rd {{International Conference}} on {{Software
  Engineering}} ({{ICSE}})}, May 2021, pp. 1161--1173.

\bibitem{xiaLessTrainingMore2022a}
C.~S. Xia and L.~Zhang, ``Less training, more repairing please: Revisiting
  automated program repair via zero-shot learning,'' in \emph{Proceedings of
  the 30th {{ACM Joint European Software Engineering Conference}} and
  {{Symposium}} on the {{Foundations}} of {{Software Engineering}}}, ser.
  {{ESEC}}/{{FSE}} 2022.\hskip 1em plus 0.5em minus 0.4em\relax ACM Press, Nov.
  2022, pp. 959--971.

\bibitem{zhuTareTypeAwareNeural2023}
Q.~Zhu, Z.~Sun, W.~Zhang, Y.~Xiong, and L.~Zhang, ``Tare: {{Type-Aware Neural
  Program Repair}},'' in \emph{Proceedings of the 45th {{International
  Conference}} on {{Software Engineering}}}, ser. {{ICSE}} '23.\hskip 1em plus
  0.5em minus 0.4em\relax IEEE Press, Jul. 2023, pp. 1443--1445.

\bibitem{wangCodeT5IdentifierawareUnified2021}
Y.~Wang, W.~Wang, S.~Joty, and S.~C.~H. Hoi, ``{{CodeT5}}: {{Identifier-aware
  Unified Pre-trained Encoder-Decoder Models}} for {{Code Understanding}} and
  {{Generation}},'' Sep. 2021.

\bibitem{li2023starcoder}
R.~Li, L.~B. Allal, Y.~Zi, N.~Muennighoff, D.~Kocetkov, C.~Mou, M.~Marone,
  C.~Akiki, J.~Li, J.~Chim \emph{et~al.}, ``Starcoder: may the source be with
  you!'' \emph{arXiv preprint arXiv:2305.06161}, 2023.

\bibitem{starEncoder}
``Starencoder model repository on huggingface,''
  https://huggingface.co/bigcode/starencoder, accessed: 2024-07-19.

\bibitem{berabi2021tfix}
B.~Berabi, J.~He, V.~Raychev, and M.~Vechev, ``Tfix: Learning to fix coding
  errors with a text-to-text transformer,'' in \emph{International Conference
  on Machine Learning}.\hskip 1em plus 0.5em minus 0.4em\relax PMLR, 2021, pp.
  780--791.

\bibitem{zhangCoditT5PretrainingSource2023}
J.~Zhang, S.~Panthaplackel, P.~Nie, J.~J. Li, and M.~Gligoric, ``{{CoditT5}}:
  {{Pretraining}} for {{Source Code}} and {{Natural Language Editing}},'' in
  \emph{Proceedings of the 37th {{IEEE}}/{{ACM International Conference}} on
  {{Automated Software Engineering}}}, ser. {{ASE}} '22.\hskip 1em plus 0.5em
  minus 0.4em\relax ACM Press, Jan. 2023, pp. 1--12.

\bibitem{roziereCodeLlamaOpen2024}
B.~Rozi{\`e}re, J.~Gehring, F.~Gloeckle, S.~Sootla, I.~Gat, X.~E. Tan, Y.~Adi,
  J.~Liu, R.~Sauvestre, T.~Remez, J.~Rapin, A.~Kozhevnikov, I.~Evtimov,
  J.~Bitton, M.~Bhatt, C.~C. Ferrer, A.~Grattafiori, W.~Xiong, A.~D{\'e}fossez,
  J.~Copet, F.~Azhar, H.~Touvron, L.~Martin, N.~Usunier, T.~Scialom, and
  G.~Synnaeve, ``Code {{Llama}}: {{Open Foundation Models}} for {{Code}},''
  Jan. 2024.

\bibitem{al2024traces}
A.~Al-Kaswan, M.~Izadi, and A.~Van~Deursen, ``Traces of memorisation in large
  language models for code,'' in \emph{Proceedings of the IEEE/ACM 46th
  International Conference on Software Engineering}, 2024, pp. 1--12.

\bibitem{sallou2024breaking}
J.~Sallou, T.~Durieux, and A.~Panichella, ``Breaking the silence: the threats
  of using llms in software engineering,'' in \emph{Proceedings of the 2024
  ACM/IEEE 44th International Conference on Software Engineering: New Ideas and
  Emerging Results}, 2024, pp. 102--106.

\bibitem{karmakar2022codex}
A.~Karmakar, J.~A. Prenner, M.~D'Ambros, and R.~Robbes, ``Codex hacks
  hackerrank: Memorization issues and a framework for code synthesis
  evaluation,'' \emph{arXiv preprint arXiv:2212.02684}, 2022.

\bibitem{sutton2019bitter}
R.~Sutton, ``The bitter lesson,'' \emph{Incomplete Ideas (blog)}, vol.~13,
  no.~1, p.~38, 2019.

\bibitem{jiangMixtralExperts2024}
A.~Q. Jiang, A.~Sablayrolles, A.~Roux, A.~Mensch, B.~Savary, C.~Bamford, D.~S.
  Chaplot, D.~de~las Casas, E.~B. Hanna, F.~Bressand, G.~Lengyel, G.~Bour,
  G.~Lample, L.~R. Lavaud, L.~Saulnier, M.-A. Lachaux, P.~Stock,
  S.~Subramanian, S.~Yang, S.~Antoniak, T.~L. Scao, T.~Gervet, T.~Lavril,
  T.~Wang, T.~Lacroix, and W.~E. Sayed, ``Mixtral of {{Experts}},'' Jan. 2024.

\bibitem{anthropicClaude}
``Claude: Models overview,''
  https://docs.anthropic.com/claude/docs/models-overview, accessed: 2024-07-31.

\bibitem{reidGeminiUnlockingMultimodal2024}
{Google Gemini Team}, ``Gemini 1.5: {{Unlocking}} multimodal understanding
  across millions of tokens of context,'' Mar. 2024.

\bibitem{liuLostMiddleHow2023}
N.~F. Liu, K.~Lin, J.~Hewitt, A.~Paranjape, M.~Bevilacqua, F.~Petroni, and
  P.~Liang, ``Lost in the {{Middle}}: {{How Language Models Use Long
  Contexts}},'' Nov. 2023.

\bibitem{hsieh2024ruler}
C.-P. Hsieh, S.~Sun, S.~Kriman, S.~Acharya, D.~Rekesh, F.~Jia, and B.~Ginsburg,
  ``Ruler: What's the real context size of your long-context language models?''
  \emph{arXiv preprint arXiv:2404.06654}, 2024.

\bibitem{vaswaniAttentionAllYou2017}
\BIBentryALTinterwordspacing
A.~Vaswani, N.~Shazeer, N.~Parmar, J.~Uszkoreit, L.~Jones, A.~N. Gomez,
  L.~Kaiser, and I.~Polosukhin, ``Attention is all you need.'' [Online].
  Available: \url{http://arxiv.org/abs/1706.03762}
\BIBentrySTDinterwordspacing

\bibitem{gu2023mamba}
A.~Gu and T.~Dao, ``Mamba: Linear-time sequence modeling with selective state
  spaces,'' \emph{arXiv preprint arXiv:2312.00752}, 2023.

\end{thebibliography}

\end{document}